\documentclass[12pt]{article}
\usepackage{graphicx}
\usepackage{amsmath,amssymb}
\newcommand{\dbox}{\,\raise2pt\hbox{\fbox{\rule{2.5pt}{0pt}\rule{0pt}{2.5pt}}}\,}
\newcommand{\qed}{\,\raise0pt\hbox{\mbox{\rule{6.5pt}{6.5pt}}}}

\begin{document}
\setlength{\baselineskip}{6.5mm}

\begin{titlepage}
 \begin{LARGE}
   \vspace*{1cm}
   \begin{center}
Minimal gauge invariant and gauge fixed actions \\ for massive higher-spin fields 
\\
   \end{center}
 \end{LARGE}
  \vspace{5mm}
 \begin{center}
    Masako {\sc Asano} 
           
      \vspace{4mm}
        {\sl Faculty of Science and Technology}\\
        {\sl Seikei University}\\
        {\sl Musashino-shi, Tokyo 180-8633, Japan}\\

       \vspace{2cm}

  ABSTRACT\par
 \end{center}
 \begin{quote}
  \begin{normalsize}
Inspired by the rich structure of covariant string field theory, we propose a minimal gauge invariant action for general massive integer spin $n$ field.
The action consists of four totally symmetric tensor fields of order respectively $n$, $n-1$, $n-2$ and $n-3$, and 
is invariant under the gauge transformation represented by two also totally symmetric fields of order $n-1$ and $n-2$.
This action exactly has the same gauge structure as for the string field theory and 
we discuss general covariant gauge fixing procedure using the knowledge of string field theory.
We explicitly construct the corresponding gauge fixed action for each of general covariant gauge fixing conditions.
\end{normalsize}
\end{quote}

\end{titlepage}
\vfil\eject

\section{Introduction}
It is well-known that a massive integer spin $n$ field on $d$ dimensional flat spacetime
is represented by totally symmetric tensor field $A_{\mu_1\cdots \mu_n}$ with the following set of conditions~\cite{Fierz:1939ix, Wigner:1939cj,Bargmann:1948ck, Bekaert:2006py, Bouatta:2004kk, Rahman:2015pzl}%
\footnote{
Here, we consider a series of irreducible representations  of o$(d-1)$ algebra classified by an integer $n$. 
For $d>4$, there exist other non-trivial representations which are described by general mixed symmetric tensor fields.
}:
\begin{align}
(\Box -m^2)A_{\mu_1\cdots \mu_n} &= 0, \label{FPb-1}\\
\partial \cdot A_{\mu_1\cdots \mu_{n-1}}&= 0,\label{FPb-2}\\
A'_{\mu_1\cdots \mu_{n-2}}&=0.\label{FPb-3}
\end{align}
A Lagrangian formulation for this system has been given by Singh and Hagen \cite{Singh:1974qz}.
The field contents of the Lagrangian are given by the set of $n$ symmetric {\it traceless} tensor fields of order $n$, $n-2$, $n-3$, $\cdots$, 0 where the highest order $n$ tensor field is the original $A_{\mu_1\cdots \mu_n}$ and the other $n-1$ lower order tensor fields are auxiliary.

Our aim of the present paper is to introduce a minimal gauge invariant Lagrangian for the same system given by eqs.(\ref{FPb-1})-(\ref{FPb-3}), and apply the gauge fixing procedure to the Lagrangian according to the knowledge of the covariant string field theory. 
The gauge structure of the minimal Lagrangian is simple enough to further attempt to construct the interaction terms consistently based on the system. 
The fields contents of the Lagrangian for general $n$ are the set of only four {\it unconstrained} symmetric tensor fields $A$, $h$, $D$, and $\phi$ of order respectively $n$, $n-1$, $n-2$, and $n-3$. 
The Lagrangian is invariant under the gauge transformation with respect to also unconstrained two gauge parameter fields $\lambda$ and $\zeta$ of order respectively $n-1$ and $n-2$.

We should notice that there have been proposed some other Lagrangian formulations for the system eqs.(\ref{FPb-1})-(\ref{FPb-3}) within the developments of higher-spin field theories, {\it e.g.}, \cite{West:1986tw, Pashnev:1989gm, Pashnev:1997rm, Fotopoulos:2008ka, Buchbinder:2005ua, Ponomarev:2010st, Buchbinder:2008ss, Francia:2007ee}.
Among them, the characteristic feature of our minimal action is that the gauge structure respects the one for the covariant string field theory, other than the simplicity of the field contents%
\footnote{We notice during preparation of this paper that the Lagrangian given in ref.\cite{Pashnev:1989gm} (and reviewed in ref.\cite{Fotopoulos:2008ka}) also consists of four unconstrained symmetric tensor fields. 
The Lagrangian should essentially be equivalent to our Lagrangian after performing a suitable field redefinition.
}.
Indeed, the action for massive spin $n$ ($n>1$) field should be extracted from the level $n$ part of the open string field theory action, though it is difficult to perform it directly because of the technical reason except for lower $n$.
The massless limit of the action is divided into the two independent actions represented respectively by the sets of two fields $(A,D)$ and $(h,\phi)$. Each of these actions is equivalent to the one known as the `triplet' action~\cite{Asano:2012qn,Bengtsson:1986ys, Francia:2002pt, Sagnotti:2003qa}, which is obtained by taking the tensionless limit of the open string field theory or by the massless part of the extended string field theory~\cite{Asano:2013rka}.

This paper is organized as follows.
In the next section~2, we propose a gauge invariant Lagrangian for massive totally symmetric field of order $n$ with three additional fields  without preparation. 
After discussing the gauge structure of the Lagrangian in detail, we show that this Lagrangian exactly represent the massive spin $n$ field step by step, by explicitly demonstrating that the fields equations (\ref{FPb-1})-(\ref{FPb-3}) are obtained from the proposed action if we fix the gauge degrees of freedom suitably. The main part of the proof is collected in Appendix~B.
We also discuss the massless limit of the Lagrangian and the relation with the triplet Lagrangian.
In section~3, we discuss covariant gauge fixing procedure for the action and give BRST invariant gauge-fixed action corresponding to each of the wide class of covariant gauge fixing conditions.
In section~4, we summarize the results and give discussions including the possible generalization and application of our result.
We collect the notations and conventions in Appendix~A.

\section{Gauge invariant Lagrangian for massive higher spin symmetric tensor field}
We first present the main result of this paper: 
We give a minimal gauge invariant Lagrangian ${\cal L}_{n}$ representing the spin $n$ field of mass $m$ in $d$-dimensional flat spacetime.
Field contents we prepare for ${\cal L}_{n}$ are four unconstrained totally symmetric tensor fields of order $n$, $n-1$, $n-2$ and $n-3$, which we write respectively as:
\begin{equation}
A_{\mu_1\mu_2\cdots\mu_{n}}, \quad h_{\mu_1\mu_2\cdots\mu_{n-1}},\quad D_{\mu_1\mu_2\cdots\mu_{n-2}},\quad \phi_{\mu_1\mu_2\cdots\mu_{n-3}}. 
\end{equation}
With this set of fields, we can construct the following quadratic Lagrangian ${\cal L}_{n}$ by assuming that the coefficients of the terms $A_{\mu_1\mu_2\cdots\mu_{n}} \Box A^{\mu_1\mu_2\cdots\mu_{n}} $ and $h_{\mu_1\mu_2\cdots\mu_{n-1}} \Box h^{\mu_1\mu_2\cdots\mu_{n-1}} $ are both non-zero and that the Lagrangian in invariant under the gauge transformations of the form eqs.(\ref{gtA})-(\ref{gtD}) and $\delta \phi=\partial\cdot  \zeta +c \partial \zeta' $.
The explicit form of the Lagrangian ${\cal L}_{n}={\cal L}_{n}(A, D, h,\phi) $ we propose is 
\begin{multline}
{\cal L}_{n}(A, D, h,\phi) 
=  {\cal L}^{AD}_{n,m^2= 0}(A,D) 
-\sum_{k=0}^{\lfloor \frac{n}{2} \rfloor} \frac{n! (1-2 k)}{2(2k)! (n-2 k)!} (m A^{[k]} - (\partial h)^{[k]})(m A^{[k]} - (\partial h)^{[k]})\\
-\sum_{k=0}^{\lfloor\frac{n-2}{2}\rfloor} \frac{3(n-2)(n-1)!}{n(2k+1)! (n-2k-2)!}
\left\{
m(D^{[k]}-n A^{[k+1]}) +n (\partial h)^{[k+1]} -\partial\cdot h^{[k]}
\right\}
 (\partial \phi)^{[k]}
\\
 -\sum_{k=0}^{\lfloor\frac{n-3}{2}\rfloor} 
\frac{9 (n-1)! (k+1)}{n(2k+3)!(n-2k-3)!}
  \left\{
2(k+2)\phi^{[k]}\Box\phi^{[k]} 
\right.
\\
\hspace*{6cm}
\left.
-(n-2k-3) \left( \partial\cdot \phi^{[k]} +2(n-2)   (\partial \phi)^{[k+1]}  \right) \partial\cdot \phi^{[k]} 
\right\}
\\
+m^2 \sum_{k=0}^{\lfloor\frac{n-3}{2}\rfloor} \frac{9 (n-1)! (k+1)}{n (2k+3)! (n-2k-3)!} (\phi^{[k]})^2
\label{LAhDphin}
\end{multline}
where
\begin{multline}
{\cal L}^{AD}_{n,m^2= 0}(A,D)  = \sum_{k=0}^{\lfloor\frac{n}{2}\rfloor} \frac{n!}{2(2k)!(n-2k)!} 
\left\{
\left(A^{[k]}-\frac{2k}{n} D^{[k-1]}  \right)  \Box \left(A^{[k]}-\frac{2k}{n} D^{[k-1]}  \right) 
\right.
\\
\left.
\hspace*{6cm}+(n-2k) \left( \partial\cdot\left(A^{[k]}-\frac{2k}{n} D^{[k-1]}  \right) \right)^2
\right\} 
\\
+\sum_{k=0}^{\lfloor\frac{n-2}{2}\rfloor} 
\frac{(n-1)!}{(2k)!(n-2k-2)!}  \left\{
 D^{[k]} \left(\partial\cdot\partial\cdot\left(A^{[k]}-\frac{2k}{n} D^{[k-1]}  \right)  \right) 
\right.
\\
\left.
-\frac{1}{n} D^{[k]}\Box D^{[k]} +\frac{n-2k-2}{2n} \left(\partial\cdot  D^{[k]}   \right)^2 
\right\} .
\label{LADm0}
\end{multline}
Here, $\Box =\partial_{\mu}\partial^\mu = -(\partial_0)^2 + (\partial_1)^2+\cdots+ (\partial_{d-1})^2$ and $\lfloor x\rfloor=$floor$(x)$ gives the greatest integer less than or equal to $x$. 
We have used the abbreviated notations of fields and derivatives in the above equations and omitted the spacetime indices. 
To restore those indices, we use the following rules.
\begin{align}
A^2 &\;=\; A_{\mu_1\mu_2\cdots\mu_{n}}A^{\mu_1\mu_2\cdots\mu_{n}}, 
\\
A^{[k]} &\;=\; A^{[k]}{}_{\mu_1 \cdots\mu_{n-2k}} 
=  A_{\mu_1\mu_2\cdots\mu_{n}}
\eta^{\mu_{n-2k+1}\mu_{n-2k+2}} \cdots \eta^{\mu_{n-1}\mu_{n}  },
\\
\partial \cdot A &\;=\; \partial^{\nu} A_{\nu\mu_1 \cdots\mu_{n-1}},
\\
(\partial A) &\;=\;\partial_{(\mu_1} A_{\mu_2\cdots\mu_{n+1})} = 
\frac{1}{n+1} \sum_{i=1}^{n+1} \partial_{\mu_i} A_{\mu_{i+1}\cdots\mu_{i+n}}, \qquad (i \sim i+n).
\end{align}
See Appendix~A for details of notations and the useful relations.
The Lagrangian ${\cal L}_{n}$ that we now just propose is constructed so that it is invariant under the following gauge transformations (up to total derivatives)
\begin{align}
\delta A &= \partial \lambda, \label{gtA}
\\
\delta h &= (n-1) \partial \zeta +m \lambda,\label{gth}
\\
\delta D &= \partial\cdot  \lambda -m \zeta,\label{gtD}
\\
\delta \phi &=\partial\cdot  \zeta +\frac{n-3}{3} \partial \zeta' \,.
\label{gtphi}
\end{align}
Here, $\lambda(=\lambda_{(\mu_1\cdots\mu_{n-1})})$ and $\zeta(=\zeta_{(\mu_1\cdots\mu_{n-2})})$ are 
unconstrained totally symmetric gauge parameter fields of order respectively $n-1$ and $n-2$.
We have also used the abbreviated notations. (Note that $\zeta'=\zeta^{[1]}$.)

We now show that the physical degrees of freedom of the system given by ${\cal L}_{n}$ exactly coincides with that of the massive spin $n$ field. 
First, note that the total number of the degrees of freedom of four fields $A$, $h$, $D$ and $\phi$ is 
\begin{equation}
{\small \left(\begin{array}{c} d+n-1\\ n \end{array}\right) +  \left(\begin{array}{c} d+n-2\\ n-1 \end{array}\right)
+\left(\begin{array}{c} d+n-3\\ n-2 \end{array}\right)+\left(\begin{array}{c} d+n-4\\ n-3 \end{array}\right) }
\label{dof4},
\end{equation}
and that of two gauge parameter fields $\lambda$ and $\zeta$ is 
\begin{equation}
{\small \left(\begin{array}{c} d+n-2\\ n-1 \end{array}\right)
+\left(\begin{array}{c} d+n-3\\ n-2 \end{array}\right)}.
\label{dofgauge}
\end{equation}
By noticing that all the gauge transformations given above are independent and that 
there is on-shell residual gauge invariance corresponding to the zero-norm (null) degrees of freedom, 
the number of physical degrees of freedom of the system ${\cal L}_n$ 
is obtained by the calculation (\ref{dof4}) $-\;2\;\times$ (\ref{dofgauge}) as 
\begin{equation}
 {\small \left(\begin{array}{c} d+n-3\\ n \end{array}\right)+ \left(\begin{array}{c} d+n-4\\ n-1 \end{array}\right) }.
\end{equation}
We see that this number exactly coincides with the degrees of freedom of the spin $n$ field $A_{(\mu_1\cdots\mu_n)}$ satisfying eqs.(\ref{FPb-1})-(\ref{FPb-3}).
We can in fact prove that the equations of motion obtained from the Lagrangian ${\cal L}_{n}(A, D, h,\phi)$ are equivalent to the set of equations (\ref{FPb-1})-(\ref{FPb-3}) after suitably fixing the gauge symmetry.
The proof is given in Appendix~B. 
With this result, we confirm that our Lagrangian ${\cal L}_n$ describes the massive spin $n$ field.

We explicitly see the structure of the Lagrangian ${\cal L}_n$ and the gauge transformation for lower $n$ as examples.
For $n=1$, they are given by 
\begin{equation}
{\cal L}_{n=1}(A_\mu, h)= -\frac{1}{4} F_{\mu\nu}F^{\mu\nu} -\frac{1}{2} (m A_\mu -\partial_\mu h)^2,
\end{equation}
and 
\begin{equation}
\delta A_\mu =\partial_\mu\lambda,\quad \delta h= m\lambda.
\end{equation}
This is the St\"{u}ckelberg Lagrangian for massive gauge field $A_\mu$ as expected.
For $n=2$, the field contents are $A_{\mu\nu},h_\mu$, and $ D$,  and the Lagrangian is given by
\begin{multline}
{\cal L}_{n=2}(A_{\mu\nu}, h_\mu, D) = \frac{1}{2}A\Box A + (\partial\cdot A )^2 +\frac{1}{2} A'\Box A' - A' \Box D + D \partial\cdot\partial\cdot A 
\\
+ \frac{1}{2}m^2(A'A'- AA) -\frac{1}{8}(\partial_\mu h_\nu -  \partial_\nu h_\mu)^2 
-m (h \partial\cdot A + A' \partial\cdot h ).
\end{multline}
This is invariant under the gauge transformation 
\begin{equation}
\delta A_{\mu\nu} =\partial_{(\mu}\lambda_{\nu)},
\quad \delta D= \partial^\mu \lambda_\mu -m \zeta,
\quad \delta h_\mu = m \lambda_\mu+\partial_\mu \zeta
\end{equation}
with gauge parameter fields $\lambda_\mu$ and $\zeta$.
This Lagrangian ${\cal L}_{n=2}$ is equivalent to the weak massive graviton field~\cite{Fierz:1939ix, Hinterbichler:2011tt} and can be obtained from the level $N=2$ of the open string field theory.
For $n=3$, the field contents are $A_{\mu\nu\rho},h_{\mu\nu}$, $D_\mu$, and $\phi$,  and the gauge parameter fields are 
$\lambda_{\mu\nu}$ and $\zeta_\mu$.
The  Lagrangian ${\cal L}_{n=3}$ is written as
\begin{multline}
{\cal L}_{n=3}(A_{\mu\nu\rho}, h_{\mu\nu}, D_\mu,\phi) =\frac{1}{2} A\Box A + \frac{3}{2}(\partial\cdot A )^2 +\frac{3}{2}A'\Box A'-2 A' \Box D 
\\
+ 2 D \partial\cdot\partial\cdot A + (\partial\cdot (D-A') )^2   +\frac{1}{2}(\partial\cdot A')^2 \\
- \frac{1}{2}(mA-\partial h)^2 +\frac{3}{2} (mA'-(\partial h)')^2 -2 \left(m(D-3A') +\partial h' +\partial\cdot h  \right) (\partial \phi) 
\\ -4\phi \Box \phi +m^2 \phi^2,
\end{multline}
which is invariant under the gauge transformation
\begin{equation}
\delta A_{\mu\nu\rho}=\partial_{(\mu}\lambda_{\nu\rho)},
\quad
\delta D_{\mu}=\partial^\nu \lambda_{\mu\nu} -m \zeta_\mu ,
\quad
\delta h_{\mu\nu}=  m \lambda_{\mu\nu} + 2 \partial_{(\mu} \zeta_{\nu)} ,
\quad
\delta \phi=\partial^\nu \zeta_{\nu}
.
\end{equation}
Note that this Lagrangian ${\cal L}_{n=3}$ is equivalent to the one obtained from the level $N=3$ part of the open string field theory~\cite{Asano:2012qn}.
The Lagrangian ${\cal L}_{n}$ for general $n>3$ should also be directly obtained from the level $N=n$ action of the open string field theory although we have not confirmed this explicitly since the string field theory action for higher level is very complicated.

The relation of the proposed Lagrangian  ${\cal L}_{n}$ to the string field theory can be seen explicitly by taking the massless limit.
If we take $m^2=0$, the Lagrangian is divided into two independent parts, the $(A,D)$ and the $(h,\phi)$ parts, as 
\begin{equation}
{\cal L}_{n}(A, D, h,\phi)|_{m^2=0} =  {\cal L}^{AD}_{n,m^2= 0}(A,D)  +  {\cal L}^{h\phi}_{n-1,m^2 = 0}\left(h,\phi\right)
\end{equation}
where ${\cal L}^{AD}_{n,m^2= 0}(A,D) $ is given in eq.(\ref{LADm0}) and ${\cal L}^{h\phi}_{n-1,m^2 = 0}\left(h,\phi\right)$ can be read from eq.(\ref{LAhDphin}).
We see from eqs.(\ref{gtA})-(\ref{gtphi}) that ${\cal L}^{AD}_{n,m^2= 0}(A,D) $  and ${\cal L}^{h\phi}_{n-1,m^2 = 0}\left(h,\phi\right)$ are invariant under the gauge transformation 
\begin{equation}
   \left\{
    \begin{aligned}
         \delta A &= \partial \lambda \\
       \delta D &= \partial\cdot  \lambda
    \end{aligned}
  \right.
\label{gaugetrAD}
\end{equation}
and
\begin{equation}
\left\{
    \begin{aligned}
        \delta h &= (n-1) \partial \zeta  \\
\delta \phi &=\partial\cdot  \zeta +\frac{n-3}{3} \partial \zeta' 
    \end{aligned}
    \right.
\end{equation}
respectively.

On the other hand, we know that the tensionless limit of the string field theory gives the following Lagrangian for totally symmetric tensor fields $A$ (order $n$) and $D$ (order $n-2$)~\cite{Asano:2012qn,Bengtsson:1986ys, Francia:2002pt, Sagnotti:2003qa}:%
\footnote{
This  Lagrangian ${\cal L}_{0,n}^{\rm string}$ is simply obtained from the massless part of the order $n$ extended string field theory developed in ref.~\cite{Asano:2013rka}.}
\begin{equation}
{\cal L}_{0,n}^{\rm string}(A,D) = \frac{1}{2}A\Box A +\frac{n}{2}(\partial\cdot A)^2+(n-1)D(\partial\cdot\partial\cdot A)
+\frac{(n-1)(n-2)}{2n}(\partial\cdot D)^2 -\frac{n-1}{n} D\Box D.
\end{equation}
This Lagrangian ${\cal L}_{0,n}^{\rm string}(A,D)$ is invariant under the same gauge transformation (\ref{gaugetrAD}) as for $ {\cal L}^{AD}_{n,m^2= 0}(A,D)$.
In fact, we can show that these two Lagrangians are equivalent to each other.
This is explicitly seen as follows.
First, note that for both systems  ${\cal L}_{0,n}^{\rm string}$ and  $ {\cal L}^{AD}_{n,m^2= 0}(A,D)$, as long as we consider off-shell fields, we can set
\begin{equation}
n\partial\cdot A- (n-1) \partial D = 0
\label{m0const}
\end{equation}
by a certain gauge transformation with the choice of suitable gauge parameter field $\lambda$ since 
\begin{equation}
\delta \left( n\partial\cdot A- (n-1) \partial D \right) = \Box\lambda
\end{equation}
holds.
Under the gauge condition (\ref{m0const}), equations of motion for both Lagrangians are reduced to the same form 
\begin{equation}
\Box A =\Box D =0.
\label{m0eomAD}
\end{equation}
This confirms that the physical systems represented by  $ {\cal L}^{AD}_{n,m^2= 0}(A,D)$ and ${\cal L}_{0,n}^{\rm string}(A,D)$ are equivalent to each other.
Note that the physical degrees of freedom can be identified by the equations (\ref{m0const}) and (\ref{m0eomAD}) after taking into account the residual on-shell gauge degrees of freedom.
The number of the physical degrees of freedom is given by $\large({\scriptsize \begin{array}{c} d+n-3\\ n \end{array}}\large)$.
Similarly, we can show that the physical system obtained from the Lagrangian ${\cal L}^{h\phi}_{n-1,m^2 = 0}\left(h,\phi\right)$ is equivalent to that from ${\cal L}_{0,n-1}^{\rm string}(\frac{h}{n-1},3 \phi-h' )$.

We finally discuss the uniqueness of the Lagrangian ${\cal L}_{n}(A, D, h,\phi)$ under the assumption of the gauge transformations eqs.(\ref{gtA})-(\ref{gtphi}).
Let us assume a quadratic Lagrangian of the form:
\begin{equation}
{\cal L}(A, D, h,\phi) = \tilde{\cal L}^{AD}_{n,m^2= 0}(A,D) + \tilde{\cal L}^{h\phi}_{n-1,m^2 = 0}\left(h,\phi \right) +{\cal L}_{m}(A,D,h,\phi) 
\label{Lass}
\end{equation}
where $\tilde{\cal L}^{AD}_{n,m^2= 0}(A,D)$ and $\tilde{\cal L}^{h\phi}_{n-1,m^2 = 0}\left(h,\phi\right)$ are assumed to have the same physical systems as ${\cal L}_{0,n}^{\rm string}(A,D)$ and ${\cal L}_{0,n-1}^{\rm string}(\frac{h}{n-1}, 3 \phi-h' )$ respectively, 
and ${\cal L}_{m}(A,D,h,\phi)$ is an interaction term between $(A, D)$ and $(h,\phi)$ fields which vanishes for $m^2=0$.
Then, we are lead to the conclusion that the Lagrangian ${\cal L}(A, D, h,\phi)$ must coincide with ${\cal L}_{n}(A, D, h,\phi)$ under the condition that the Lagrangian is invariant under the gauge transformation eqs.(\ref{gtA})-(\ref{gtphi}).
That is, the Lagrangian ${\cal L}_{n}(A, D, h,\phi)$ is uniquely obtained by the request of the gauge invariance.
As for the choice of the gauge transformation, we may notice that the gauge transformation $\delta\phi$ given by eq.(\ref{gtphi}) is not the simplest choice.
However, if we instead choose the simpler form as $\delta\phi = \partial\cdot\zeta$,  
we cannot obtain any nontrivial gauge invariant Lagrangian of the form given by eq.(\ref{Lass}).
In fact, if we assume $\delta\phi = \partial\cdot\zeta + c \partial \zeta'$ with an arbitrary constant $c$ instead of eq.(\ref{gtphi}) and repeat the same argument, we reach the same Lagrangian ${\cal L}_{n}(A, D, h,\phi)$ with $c=\frac{n-3}{3}$.

\section{Gauge fixing procedure and the gauge fixed action}
We now discuss covariant gauge fixing procedure for the gauge invariant Lagrangian ${\cal L}_{n}$ proposed in the previous section.
We can identify a set of appropriate gauge fixing conditions which is parametrized by $n$ real parameters. 
This is seen by explicitly constructing a consistent BRST invariant gauge fixed action for each of the conditions.
In order to proceed with the discussion, we first define the following two totally symmetric tensor fields $f_\lambda$ (order $n-1$) and $f_\zeta $ (order $n-2$) as
\begin{align}
f_\lambda &\equiv n\partial\cdot A-(n-1)\partial D -m h ,
\label{flambda}
\\
f_\zeta &\equiv \partial\cdot h +(n-2) \partial (h' -3 \phi) -nmA' +mD.
\label{fzeta}
\end{align}
We see from eqs.(\ref{gtA})-(\ref{gtphi}) that the gauge transformation of these fields takes the following specific form  
\begin{equation}
\delta f_\lambda = (\Box-m^2)\lambda,
\qquad
\delta f_\zeta= (\Box-m^2)\zeta.
\label{gaugetrflz}
\end{equation}
When the fields are off-shell $(\Box-m^2\ne 0)$, we can gauge transform to set $ f_\lambda = f_\zeta=0 $ by adjusting the gauge parameter fields $\lambda$ and $\zeta$.
Since there is no additional gauge degrees of freedom within the condition $ f_\lambda = f_\zeta=0 $, this gives an appropriate gauge fixing condition.
In fact, this particular condition corresponds to the simple gauge known as Landau gauge. 
Other than this choice of gauge, we can choose many other consistent covariant gauge fixing conditions. 
We identify a class of consistent gauge fixing conditions by explicitly constructing an appropriate gauge fixed action for each of these gauge fixing conditions. 
General form of such gauge fixed action is given by
\begin{equation}
S^{n}_{\{\alpha_1,\alpha_2,a_i, b_i\}}=\int d^d x \left( {\cal L}_{n} + {\cal L}_{{\rm gh+g.f.},\{\alpha_1,\alpha_2,a_i, b_i\}}^{n} \right)
\label{gfaction}
\end{equation}
where ${\cal L}_{{\rm gh+g.f.},\{\alpha_1,\alpha_2,a_i, b_j\}}^{n}  $ represents the (anti-)ghosts and the gauge fixing terms with $n$ real parameters
 $\alpha_1$, $\alpha_2$, $a_1,\cdots,a_{\lfloor \frac{n-1}{2}\rfloor }$,  $b_1,\cdots,b_{\lfloor \frac{n-2}{2} \rfloor}$  which distinguish the different gauge conditions.
 (Note that $\lfloor \frac{n-1}{2}\rfloor +\lfloor \frac{n-2}{2}\rfloor =n-2$.)
 The explicit form of ${\cal L}_{{\rm gh+g.f.},\{\alpha_1,\alpha_2,a_i, b_j\}}^{n}  $ is given by
\begin{multline}
{\cal L}_{{\rm gh+g.f.},\{\alpha_1,\alpha_2,a_i, b_j\}}^{n}   = 
\alpha_1 \beta_\lambda{}^2 +\alpha_2 \beta_{\zeta}{}^2
+
\sum_{k=0}^{\lfloor \frac{n-1}{2}\rfloor }  a_k \beta_\lambda^{[k]} f_{\lambda}^{[k]}
+ 
\sum_{k=0}^{\lfloor \frac{n-2}{2}\rfloor }  b_k \beta_\zeta^{[k]} f_{\zeta}^{[k]}
\\
+ i 
\sum_{k=0}^{\lfloor \frac{n-1}{2}\rfloor } a_k \bar{\gamma}_\lambda^{[k]}(\Box-m^2)\gamma_{\lambda}^{[k]} 
+i 
\sum_{k=0}^{\lfloor \frac{n-2}{2}\rfloor }  b_k \bar{\gamma}_\zeta^{[k]}(\Box-m^2)\gamma_{\zeta}^{[k]} 
\label{gfixterms}
\end{multline}
where $a_0=b_0=1$. 
We have introduced several new fields:  $\beta_\lambda(=\beta_{\lambda\;(\mu_1\cdots\mu_{n-1} )})$ and  $\beta_\zeta(=\beta_{\zeta\;(\mu_1\cdots\mu_{n-2} )})$, which correspond to Nakanishi-Lautrup (NL) fields, 
 ghost fields $\gamma_\lambda(=\gamma_{\lambda\;(\mu_1\cdots\mu_{n-1} )})$,  $\gamma_\zeta(=\gamma_{\zeta\;(\mu_1\cdots\mu_{n-2} )})$,  
and anti-ghost fields $\bar{\gamma}_\lambda(=\bar{\gamma}_{\lambda\;(\mu_1\cdots\mu_{n-1} )})$,  $\bar{\gamma}_\zeta(=\bar{\gamma}_{\zeta\;(\mu_1\cdots\mu_{n-1} )})$.
Note that ghost and anti-ghost fields are Grassmann odd.
Note also that the third and the forth terms of the right-hand side of eq.(\ref{gfixterms}) can be rewritten as 
\begin{align}
\sum_{k=0}^{\lfloor \frac{n-1}{2}\rfloor }  a_k \beta_\lambda^{[k]} f_{\lambda}^{[k]}
&= \beta_\lambda\left(
f_{\lambda} +a_1 f_{\lambda}'\eta +a_2 f_{\lambda}''\eta\eta +\cdots + a_{\lfloor \frac{n-1}{2}\rfloor } f_{\lambda}^{[\lfloor \frac{n-1}{2}\rfloor ]}\eta\eta{\tiny \cdots}\eta
\right),
\\
\sum_{k=0}^{\lfloor \frac{n-2}{2} \rfloor }  b_k \beta_\zeta^{[k]} f_{\zeta}^{[k]}
&= \beta_\zeta\left(
f_{\zeta} +a_1 f_{\zeta}'\eta +a_2 f_{\zeta}''\eta\eta +\cdots + b_{\lfloor \frac{n-2}{2}\rfloor } f_{\zeta}^{[\lfloor \frac{n-2}{2}\rfloor ]}\eta\eta{\tiny \cdots}\eta
\right)
\end{align}
in the abbreviated notation given in Appendix~A.
In order that ${\cal L}_{{\rm gh+g.f.},\{\alpha_1,\alpha_2,a_i, b_j\}}^{n}  $ gives the gauge fixing terms for an appropriate gauge fixing condition,
$a_1,\cdots,a_{\lfloor \frac{n-1}{2}\rfloor }$,  $b_1,\cdots,b_{\lfloor \frac{n-2}{2} \rfloor }$ should satisfy the additional conditions: 
The parameters should be taken so that the conditions 
\begin{equation}
f_\lambda=0 ,\quad  \mbox{and}  
\quad
f_\zeta=0
\label{f0f0}
\end{equation}
are definitely lead from the assumption 
\begin{equation}
\sum_{k=0}^{\lfloor \frac{n-1}{2}\rfloor }  a_k  f_{\lambda}^{[k]} =0 ,\quad  \mbox{and}  
\quad
\sum_{k=0}^{\lfloor \frac{n-2}{2}\rfloor }  b_k f_{\zeta}^{[k]}=0 
\qquad (a_0=b_0=1).
\end{equation}
Also we can choose $\alpha_1 \ge 0$ and $\alpha_2 \ge  0$ without loss of generality. 
If we choose a set of parameters $\{ \alpha_1,\alpha_2,a_i, b_j \}$ satisfying these conditions, 
the action $S^{n}_{\{\alpha_1,\alpha_2,a_i, b_i\}}$ gives an appropriate gauge fixed action since it does not have any remaining gauge degrees of freedom and gains BRST and anti-BRST invariance instead.  
The BRST transformation is given by  
\begin{align}
\delta_B A &= -i \xi\partial \gamma_\lambda, 
\\
\delta_B h &= -i \xi  \left( (n-1) \partial \gamma_\zeta +m \gamma_\lambda \right),
\\
\delta_B D &= -i \xi \left( \partial\cdot  \gamma_\lambda -m \gamma_\zeta\right),
\\
\delta_B \phi &=-i \xi  \left(\partial\cdot  \gamma_\zeta +\frac{n-3}{3} \partial \gamma_\zeta' \right),
\\
\delta_B \bar{\gamma}_\lambda & =\xi \beta_\lambda,
\\
\delta_B \bar{\gamma}_\zeta & =\xi \beta_\zeta,
\end{align}
and $\delta_B {\gamma}_\lambda =\delta_B {\gamma}_\zeta=0$
where $\xi$ is a Grassmann odd parameter.
Also, the anti-BRST transformation is given by
\begin{align}
\tilde{\delta}_B A &= i \tilde{\xi} \partial \bar{\gamma}_\lambda, 
\\
\tilde{\delta}_B h &= i \tilde{\xi} \left( (n-1) \partial \bar{\gamma}_\zeta +m \bar{\gamma}_\lambda \right),
\\
\tilde{\delta}_B D &= i \tilde{\xi}\left( \partial\cdot  \bar{\gamma}_\lambda -m \bar{\gamma}_\zeta\right),
\\
\tilde{\delta}_B \phi &=i \tilde{\xi} \left(\partial\cdot  \bar{\gamma}_\zeta +\frac{n-3}{3} \partial \bar{\gamma}_\zeta' \right),
\\
\tilde{\delta}_B {\gamma}_\lambda & =\tilde{\xi} \beta_\lambda,
\\
\tilde{\delta}_B {\gamma}_\zeta & =\tilde{\xi} \beta_\zeta,
\end{align}
and $\tilde{\delta}_B \bar{\gamma}_\lambda =\tilde{\delta}_B \bar{\gamma}_\zeta=0$ where $\tilde{\xi}$ is a Grassmann odd parameter.
These two transformations commute with each other 
\begin{equation}
\left[  \delta_B, \tilde{\delta}_B \right] =0 .
\end{equation}
Note that we have taken $\delta_B$ and $\tilde{\delta}_B$ as Grassmann even quantities since we have included Grassmann odd parameters $\xi$ and $\tilde{\xi}$ in the definition of the transformations.
This structure is exactly the same as that for the minimal gauge fixed action constructed from the covariant string field theory~\cite{Bochicchio:1986bd, Bochicchio:1986zj, Thorn:1986qj, Asano:2006hk}.

For any $n$, the choice $\alpha_1=\alpha_2=0$ corresponds to the Landau gauge since the condition (\ref{f0f0}) is obtained from the gauge fixing terms.
Furthermore, we see the explicit form of the gauge fixed Lagrangian for $n=1$ and $n=3$ as examples.
For $n=1$, there is only one gauge parameter $\alpha$ and the gauge fixing terms are given by
\begin{equation}
{\cal L}_{{\rm gh+g.f.},\{\alpha\}}^{n=1}   = \alpha \beta^2  
+\beta( \partial\cdot A-mh)
+i \bar{\gamma}(\Box-m^2)\gamma
\end{equation}
where $\beta$ is the NL field, $\gamma$ and $\bar{\gamma}$ are ghost and anti-ghost fields. 
This coincides with the gauge fixing terms for well-known one-parameter family of covariant gauges for gauge field $A_\mu$.
The Feynman gauge and the Landau gauge correspond to $\alpha=\frac{1}{2}$ and $\alpha=0$ respectively.
The case of $n=3$ gives a more complicated example. 
In this case, there are three parameters 
$\{\alpha_1, \alpha_2,a_1\}$ which satisfy $\alpha_1\ge 0$, $\alpha_2\ge 0$ and $a_1\ne -\frac{1}{d}$.
The condition for $a_1$ comes from the fact that 
the equation $f_\lambda=0$ cannot be lead from the condition $f_{\lambda\,(\mu_1\mu_2)} + a_1 f_{\lambda}'\,\eta_{\mu_1\mu_2}=0 $ for $a_1=-\frac{1}{d}$. 
Besides, if we assume $a_1\ne -\frac{1}{d}$, $f_\lambda=0$ is lead from the same condition.
The explicit form of the gauge fixing terms for ${\cal L}_{n=3}$ are given by
\begin{multline}
{\cal L}_{{\rm gh+g.f.},\{\alpha_1,\alpha_2,a_1\}}^{n=3}   = 
\alpha_1 \beta_\lambda{}^2 +\alpha_2 \beta_{\zeta}{}^2
+
\beta_\lambda f_{\lambda} +a_1 \beta'_\lambda f'_{\lambda}
+ 
\beta_\zeta f_{\zeta}
\\
+ i \bar{\gamma}_\lambda(\Box-m^2)\gamma_{\lambda} + i  a_1 \bar{\gamma}'_\lambda(\Box-m^2)\gamma'_{\lambda}
+i \bar{\gamma}_\zeta (\Box-m^2)\gamma_{\zeta} .
\end{multline}
The gauge condition specified by $\alpha_1=\alpha_2(=a_1)=0$ corresponds to the Landau gauge.
The Feynman gauge, where the propagators for all the fields are given by $\frac{1}{\Box-m^2}$,
is given by choosing the parameters as $(\alpha_1, \alpha_2,a_1)=(\frac{3}{2}, \frac{3}{4}, -d\pm \sqrt{d})$.

\section{Summary and discussions}
We have explicitly constructed a minimal gauge invariant Lagrangian ${\cal L}_n$ for free massive integer spin $n$ field whose gauge structure conforms to the gauge invariant action of covariant string theory.
We have also identified a set of appropriate covariant gauge fixing conditions by using the knowledge of the string field theory and have given consistent BRST and anti-BRST invariant gauge fixed action for each gauge condition.
The main results are given in eqs.(\ref{LAhDphin}), (\ref{LADm0}), (\ref{gfaction}) and  (\ref{gfixterms}).

The Lagrangian ${\cal L}_n$, which consists of four unconstrained totally symmetric tensor fields, can be rewritten as another form by introducing two additional auxiliary fields. This form of the Lagrangian thus consists of six independent tensor fields and is divided into two triplet Lagrangians when we take the massless limit.
The explicit form of the Lagrangian should be determined by assuming the gauge transformation given by eqs.(\ref{gtA})-(\ref{gtphi}) and  the one for new fields: 
\begin{equation}
\delta C = \Box \lambda +m (n-1)  \partial \zeta,
\qquad
 \delta B =\Box \zeta +m (n-2)  \partial \lambda' + m (n-1)  \partial\cdot \lambda
\end{equation}
where $C$ and $B$ are the auxiliary totally symmetric tensor fields of order $n-1$ and $n-2$ respectively. 

In the present paper, we have only dealt with the integer spin $n$ fields, which are the simplest of various massive higher spin fields.
In principle, we are able to construct a minimal Lagrangian of the same kind for any other field of any massive representation since all we need is included in the massive tower of string field theories. 
Such Lagrangian is in general assumed to represent fields of reducible representations as in the case of the triplet Lagrangian for massless totally symmetric fields.
For example, it might be straightforward to construct the minimal gauge invariant Lagrangian for fermionic spinor-tensor field $\psi^a_{\mu_1\cdots \mu_n}=\psi^a_{(\mu_1\cdots \mu_n)}$  based on the massive part of the supersrting field theory. 
The field contents we provide for such Lagrangian (for general $n>2$) should be the following six unconstrained totally symmetric spinor-tensor fields:
\begin{equation}
\psi^a _{\mu_1\cdots \mu_n},\; \chi^a _{\mu_1\cdots \mu_{n-1}}, \;\omega^a _{\mu_1\cdots \mu_{n-2}}, 
\quad
\tilde{\psi}^a _{\mu_1\cdots \mu_{n-1}},\; \tilde{\chi}^a _{\mu_1\cdots \mu_{n-2}}, \; \tilde{\omega}^a _{\mu_1\cdots \mu_{n-3}}.
\end{equation}
The explicit form of the Lagrangian for these fields should be constructed by assuming a suitable form of gauge transformations so than it is divided into two triplet Lagrangians of massless higher-spin fermionic fields~\cite{Francia:2002pt, Sagnotti:2003qa} when we take the massless limit.
After constructing the Lagrangian explicitly, we will be able to clarify the relation between such system and the massive half-integer spin $s=n+\frac{1}{2}$ field given by 
the totally symmetric spinor-tensor field $\psi^a_{\mu_1\cdots \mu_n}$ satisfying the conditions~\cite{Fierz:1939ix, Singh:1974rc}:
\begin{align}
(i/\!\!\!\partial   -m)\psi^a _{\mu_1\cdots \mu_n} &= 0,\\
\partial \cdot \psi^a_{\mu_1\cdots \mu_{n-1}}&= 0,\\
\gamma^\nu\psi^a_{\nu\mu_1\cdots \mu_{n-1}}&=0
\end{align}
with the spinor index $a=1,\cdots, 2^{\lfloor \frac{d}{2} \rfloor } $.
Also, it would be possible to construct a minimal Lagrangian for a massless or massive field of general non-trivial representation.
The investigation of the massless action of the extended string field theory \cite{Asano:2013rka, Asano:2016rxi} would be helpful for such purpose. We leave such problem for future work.

\section*{Acknowledgements}
The author would like to thank M.~Kato for discussions and comments.

\appendix 
\def\thesection{Appendix~\Alph{section}}
\renewcommand{\theequation}{\Alph{section}.\arabic{equation}}
\def\thesection{Appendix~\Alph{section}}
\setcounter{equation}{0}
\setcounter{figure}{0}
\def\thesection{Appendix~\Alph{section}}
\section{Notations and Conventions}
\label{app1}
We give the notations and conventions used in the text.
All the fields we deal with in the text are totally symmetric tensor fields and we often omit the parentheses indicating the symmetrization of the indices as $A_{\mu_1\cdots\mu_n} \equiv A_{(\mu_1\cdots\mu_n)} $. 
Note that we normalize the indices as $$(\mu\nu)\equiv\frac{1}{2} \mu\nu+ \frac{1}{2} \nu\mu.$$
We use the mostly plus metric $\eta_{\mu\nu}$ for $d$-dimensional flat spacetime on which our theory is defined. 
For the operation of derivatives with respect to spacetime coordinates $x_\nu$ on $A_{(\mu_1\cdots\mu_n)}$, 
we use the abbreviated notation 
\begin{align}
\partial \cdot A_{\mu_1\cdots\mu_{n-1}} &\equiv \partial^\nu A_{\nu\mu_1\cdots\mu_{n-1}},\\
(\partial A)_{\mu_1\cdots\mu_{n+1}} &\equiv \partial_{(\mu_1} A_{\mu_2\cdots\mu_{n+1})}.
\end{align}
We may further omit the indices and write $\partial\cdot A$ or $(\partial A)$ if possible.
We take the symmetrization of the indices of the product of two symmetric tensors $A_{\mu_1\cdots\mu_{n}}$ and $B_{\mu_1\cdots\mu_{n}}$ 
and denote the result by $(AB)$.
For example, symmetrization of the indices of the product of $A$ and two metric tensors is denoted by  
\begin{equation}
(\eta \eta A)_{\mu_1\cdots\mu_{n+4}} = \eta_{(\mu_1\mu_2}\eta_{\mu_{3}\mu_{4}}  A_{\mu_{5}\cdots\mu_{n+4})}.
\end{equation}
Note that the contraction of  two  totally symmetric tensor fields $A$ and $B$ of the same order $n$ is denoted by
$A B $: $A B = A_{\mu_1\cdots\mu_n} B^{\mu_1\cdots\mu_n}$.
Note that $AB$ gives a scalar unlike $(AB)$ gives a $2n$-th order symmetric tensor field. 
We use the prime index $'$ to represent the trace operation as 
\begin{equation}
A'_{\mu_1\cdots\mu_{n-2}}=\eta^{\nu\rho} A_{\nu\rho\mu_1\cdots\mu_{n-2}}.
\end{equation}
When we take trace operations $k$ times, we use the notation $[k]$ as  
\begin{equation}
A^{[k]}_{\mu_1\cdots\mu_{n-2k}}=\eta^{\nu_1\rho_1}\cdots\ \eta^{\nu_k\rho_k} A_{\nu_1\rho_1\cdots \nu_k\rho_k\mu_1\cdots\mu_{n-2k}},
\end{equation}
{\it i.e.}, $A^{[1]}=A'$,  $A^{[2]}=A''$ and so on. 
For example, for an $n$-th order tensor field $A=A_{\mu_1\cdots\mu_n} $ in $d$-dimensional spacetime, the following relations hold.
\begin{align}
\partial \cdot (\partial A) & =  \frac{1}{n+1} \Box A+\frac{n}{n+1} (\partial \partial A), \\
(\partial A)^{[k]} & = \frac{2k}{n+1} \partial\cdot A^{[k-1]}+\frac{n-2k+1}{n+1} \partial A^{[k]}
\\
(A\underbrace{\eta\cdots\eta}_{k})' & =  \frac{n(n-1)}{(n+2k)(n+2k-1)}A'\underbrace{\eta\cdots\eta}_{k}
+ \frac{2k(2k-2+2n+d)}{(n+2k)(n+2k-1)}A\underbrace{\eta\cdots\eta}_{k-1}
.
\end{align}

\section{Physical equivalence between ${\cal L}_{n}$ and the field equations (\ref{FPb-1})-(\ref{FPb-3})}
We prove that the equations of motion obtained from the Lagrangian ${\cal L}_{n}(A, D, h,\phi)$ given in eq.(\ref{LAhDphin}) coincide with the set of equations (\ref{FPb-1})-(\ref{FPb-3}) after suitably fixing the gauge symmetry.
We use the two fields $f^\lambda_{\mu_1\cdots\mu_{n-1}}$ and $f^\zeta_{\mu_1\cdots\mu_{n-2}}$ defined in the text: 
\begin{align}
(\ref{flambda}) & \qquad
f^\lambda = n\partial\cdot A-(n-1)\partial D -m h ,
\\
(\ref{fzeta}) & \qquad
f^\zeta = \partial\cdot h +(n-2) \partial (h' -3 \phi) -nmA' +mD.
\end{align}
From the relation (\ref{gaugetrflz}), we can completely fix the off-shell gauge symmetry by setting 
\begin{equation}
f^\lambda=f^\zeta=0.
\label{offred}
\end{equation}
Then, there remains residual on-shell gauge symmetry within the condition $f^\lambda=f^\zeta=0$ since 
$\delta f_\lambda = \delta f_\zeta =0$ for on-shell fields satisfying $(\Box-m^2) \lambda =(\Box-m^2) \zeta =0$.  
By using this residual symmetry, we can further set 
\begin{equation} 
h=D=0
\end{equation}
assuming that the fields satisfy the on-shell condition. 
Note that under the condition $h=D=0$, the gauge condition (\ref{offred}) is reduced to 
\begin{align}
& \partial\cdot A=0,
\label{eqst1}
\\
& A'=-\frac{3(n-2)}{nm} \partial\phi.
\label{eqst2}
\end{align}
We see that the equations of motion obtained from the Lagrangian ${\cal L}_n$ under the condition $h=D=0$ with  (\ref{eqst1}) and (\ref{eqst2}) can be reduced to the equations
\begin{align}
& (\Box-m^2 ) A=0,
\\
&   (\Box-m^2 ) \phi=0,
\label{eomphi}
\end{align}
By operating $\partial\cdot$ on the equation (\ref{eqst2}) and using the relation (\ref{eqst1}), we obtain $\partial\cdot(\partial\phi) =0$, which is equivalent to $\Box \phi +(n-3)\partial(\partial\cdot \phi)=0$. 
Then, by using eq.(\ref{eomphi}), we obtain
\begin{equation}
\phi=-\frac{n-3}{m^2}\partial(\partial\cdot\phi).
\label{Bindk1}
\end{equation}
Furthermore,  we have the relation
\begin{equation}
k\underbrace{\partial\cdot\partial\cdot \cdots \partial\cdot}_{k-1} \phi =-\frac{n-k-2}{m^2} \partial (\underbrace{\partial\cdot\partial\cdot \cdots \partial\cdot }_k\phi ),
\label{Brel1}
\end{equation}
which can be proven by induction (from the relation (\ref{Bindk1})).
Recalling that $\phi$ is the order $n-3$ tensor field, we see that the relation (\ref{Brel1}) for $k=n-2$ (with eq.(\ref{eomphi})) is reduced to 
\begin{equation}
(n-2) \underbrace{\partial\cdot\partial\cdot \cdots \partial\cdot}_{n-3} \phi =0 .
\end{equation}
Applying this result to the right-hand side of the relation (\ref{Brel1}) for $k=n-3$, we obtain $\underbrace{\partial\cdot\partial\cdot \cdots \partial\cdot}_{n-4} \phi =0 $. Repeating the same argument by applying $k=n-4, n-5,\cdots, 1$ for (\ref{Brel1}) in order,  we finally show $\phi=0$ and thus $A'=0$ from eq.(\ref{eqst2}).
Thus, we have shown that the equations of motion under the gauge fixing condition $h=D=0$, (\ref{eqst1}) and (\ref{eqst2}) are reduced to 
$\phi=0$ and 
\begin{align}
& (\Box-m^2 ) A=0,
\\
& \partial\cdot A=0,
\\
& A'=0.
\end{align}
This set of equations exactly coincides with the conditions (\ref{FPb-1})-(\ref{FPb-3}).
This completes the proof of the equivalence between the physical system given by our minimal gauge invariant Lagrangian ${\cal L}_n$ 
and the known field equations (\ref{FPb-1})-(\ref{FPb-3}) for massive higher-spin field.


\end{document}